\documentclass[sigconf]{acmart}

\copyrightyear{2024} 
\acmYear{2024} 
\setcopyright{rightsretained} 
\acmConference[IWSPA '24]{Proceedings of the 10th ACM International Workshop on Security and Privacy Analytics}{June 21, 2024}{Porto, Portugal}
\acmBooktitle{Proceedings of the 10th ACM International Workshop on Security and Privacy Analytics (IWSPA '24), June 21, 2024, Porto, Portugal}
\acmDOI{10.1145/3643651.3659890}
\acmISBN{979-8-4007-0556-4/24/06}

\usepackage{amsmath,amsfonts}
\usepackage{graphicx}

\usepackage{paralist}

\usepackage{csquotes}

\begin{document}
\settopmatter{printacmref=false}
\title{Legally Binding but Unfair?\\ Towards Assessing Fairness of Privacy Policies}

\author{Vincent Freiberger}
\email{freiberger@cs.uni-leipzig.de}
\affiliation{%
  \institution{Center for Scalable Data Analytics and Artificial Intelligence (ScaDS.AI)}
  \city{Dresden/Leipzig}
  \country{Germany}
}

\author{Erik Buchmann}
\email{buchmann@informatik.uni-leipzig.de}
\affiliation{%
  \institution{Center for Scalable Data Analytics and Artificial Intelligence (ScaDS.AI)}
  \city{Dresden/Leipzig}
  \country{Germany}
}

\begin{abstract}
Privacy policies are expected to inform data subjects about their data protection rights and should explain the data controller's data management practices.  Privacy policies only fulfill their purpose, if they are correctly interpreted, understood, and trusted by the data subject. 
This implies that a privacy policy is written in a \textit{fair} way, e.g., it does not use
polarizing terms, does not require a certain education, or does not assume a particular social background. 

We outline our approach to assessing fairness in privacy policies. We identify from fundamental legal sources and fairness research, how the dimensions \textit{informational fairness}, \textit{representational fairness} and \textit{ethics~/~morality} are related to privacy policies. We propose options to automatically assess policies in these fairness dimensions, based on text statistics, linguistic methods and artificial intelligence. We conduct initial experiments with German privacy policies to provide evidence that our approach is applicable. Our experiments indicate that there are issues in all three dimensions of fairness. 
This is important, as future privacy policies may be used in a corpus for legal artificial intelligence models.

\end{abstract}



\maketitle

\section{Introduction}
\label{sec:intro}

The General Data Protection Regulation (GDPR)~\cite{eu2016regulation} requires any organization, that manages personal data, to publish a privacy policy. A policy should make transparent how personally identifiable information is collected, shared, or used~\cite{Zaeem2020}. 
Privacy policies fulfill an important social task: They balance the information deficit between the data subject and holder, and create trust~\cite{starke2022fairness}. 

To ensure that privacy policies are properly understood, perceived, and accepted, they must be written in a \textit{fair} way. 
This includes multiple dimensions. 
\textit{Informational Fairness} is not just about what information is communicated, but also how it is communicated~\cite{boudjella2017non,trzepla2019,rello2017present,evans2014evaluation,aikens2008socioeconomic}. A negative example would be a policy consisting of complex legal phrases, which discriminates against people with dyslexia and non-native speakers.
\textit{Representational Fairness} is about biases towards certain groups represented in texts~\cite{Rice2019,baker-gillis-2021-sexism,Gumusel2022}. For example, a policy could discriminate against women by only using male word forms.
\textit{Ethics and Morality} are at the core of fairness~\cite{Hooker2005, Schwobel2022} and to some extent captured by legal frameworks~\cite{oecdprivacyguidelines,eu2016regulation}. A negative example would be a policy, that tries to take exclusive rights to utilize the user's personal data. 
However, many different definitions of fairness exist~\cite{verma2018fairness,mehrabi2021survey}, some of them are contradictory~\cite{Chouldechova2017Big,Defrance2023}, and to the best of our knowledge, none of them is tailored for privacy policies. 
It is also unclear, how well methods from natural language processing or artificial intelligence allow evaluating privacy policies for such issues. 
Thus, our research question is as follows: 

\textit{How can we automatically assess informational fairness, representational fairness, and ethics~/~morality of privacy policies?}


In this paper, we relate prominent definitions and concepts of fairness and bias to privacy policies. We propose approaches for automatically evaluating privacy policies, and we test this with selected policies. 
We make three contributions:

\begin{compactitem}
    \item We compare fairness definitions and related concepts for their applicability to privacy policies.
    \item We propose an approach based on readability metrics, lexical filtering, and large language models to assess informational and representational fairness as well as ethics~/~morality of privacy policies. 
    \item With a series of preliminary experiments, we assess the applicability of our approach to real-world privacy policies from the German Top-100 web shops. 

\end{compactitem}
To the best of our knowledge, we are the first to suggest an approach for investigating the fairness of privacy policies. By that, we set the foundation to shed light on the fairness of legal texts on a linguistic level which has received barely any attention from previous research. Our preliminary results give reason to go further. The aim is to provide data subjects with some much-needed transparency and avoid discrimination or unethical practices. 

\textbf{Paper structure:}
Section~\ref{sec:related} reviews related work. 
In Section~\ref{sec:statement}, we derive our problem statement from legal requirements.
The Sections~\ref{sec:informational}, \ref{sec:representational} and \ref{sec:ethics} explain how three dimensions of fairness can be automatically assessed. 
Section~\ref{sec:uses} summarizes potential applications and use cases of our approach. 
Finally, Section~\ref{sec:conclusion} concludes.

\section{Related Work}
\label{sec:related}
This section introduces fairness definitions, related machine-learning and NLP approaches, and measures for fairness-relevant aspects.

\subsection{The Concept of Fairness}
Fairness is a complex, context-dependent, and ambiguously defined concept~\cite{Schwobel2022,verma2018fairness, Hooker2005}. 
Relevant for our work are \textit{individual attitudes}, \textit{biases}, and \textit{legality, morality and ethics}~\cite{landers2022auditing}. 

\subsubsection{Individual attitudes} This includes distributive, procedural, and interactional justice perceptions~\cite{greenberg1990organizational}. We notice that justice perceptions are often used interchangeably with fairness. Others also split interactional justice into interpersonal and informational justice~\cite{Colquitt2015O}. 
Because privacy policies are equally accessible to every user, we do not need to consider allocation aspects (distributive fairness), and focus on procedural and informational fairness. 

\textbf{Procedural fairness}~\cite{doherty2012ends} means unbiased and non-ideological procedures. They should represent individuals involved, and rely on accurate information. Fair processes should respectfully treat affected individuals.
\textit{Assessing procedural fairness} can be based on how procedures \enquote{suppress bias, create consistent allocations, rely on accurate information, are correctable, represent the concerns of all recipients, and are based on moral and ethical standards}~\cite{greenberg1990organizational}. 

\textbf{Informational fairness}~\cite{Colquitt2015O, Schoeffer2022} is about clearly, consistently and reasonably explained processes and whether the information provided is suited to individuals' needs. Informational fairness addresses how complete and specific the disclosed information is and how readable and comprehensible it is to its audience. Informational fairness is closely related to transparency, which can be seen as an aspect of procedural fairness~\cite{Lee2019}.

\subsubsection{Biases} Embedded meanings in language can be biased~\cite{landers2022auditing}. A bias is a \enquote{dynamic and social and not [just] a statistical issue}~\cite{ntoutsi2020bias}. Biases impact fairness, because they can be harmful to specific groups~\cite{blodgett2020language}. A prominent bias in text corpora is the presence of stereotypes, which is addressed by representational fairness. 

\textbf{Representational fairness}~\cite{abbasi2019fairness} can be a source of harm, because language establishes power relationships and represents social identities~\cite{blodgett2020language}. A bias with stereotypical associations of a demographic group can lead to discrimination. Such a demographic or social group is called protected group. Discrimination is the unjustified difference in the treatment of individuals based on their membership in protected (sub)groups~\cite{edenberg2023disambiguating}, and can be a source of unfairness~\cite{mehrabi2021survey,wachter2021fairness}. 
Representational unfairness materializes in word embeddings as toxicity, stereotyping, or other forms of misrepresentation of protected groups, and has received much attention~\cite{caliskan2017semantics,papakyriakopoulos2020bias,Rice2019,blodgett2020language,garg2018,Gumusel2022,schroder2021evaluating} already.

\subsubsection{Morality and ethics} 
Morality and ethics play an important role in fairness~\cite{landers2022auditing,Hooker2005,Schwobel2022}. 
Privacy ethics investigates access provided to others to personal information and how much control one has regarding one's information being collected, stored, and used by others~\cite{decew1986scope}. In the context of modern technologies, it discusses complex privacy trade-offs and power relationships between the data holder and data subject~\cite{acquisti2015privacy,clifford2018data}. Ethics and law share a coordinating function and complement each other~\cite{rochel2021ethics}. It uses values and norms to interpret law. The latter addresses soft ethics which \cite{floridi2018soft} distinguishes from hard ethics which shape legislation like the GDPR and are embedded in it.

\subsection{Machine Learning and NLP Approaches}
Various approaches have been proposed to assess textual features that are correlated with certain fairness aspects. 
We introduce \textit{readability metrics}, \textit{bias metrics}
and \textit{morality classifications}. 

\subsubsection{Readability Metrics}
\label{subsec:readbl}
Related metrics exist on three levels. On the \textbf{word level}, foreign words, anglicisms, long words, abbreviations, complex words, etc., are problematic for non-native speakers~\cite{boudjella2017non}, elderly~\cite{trzepla2019}, dyslexics~\cite{rello2017present} or autistic people~\cite{evans2014evaluation}, or socioeconomically less-privileged~\cite{aikens2008socioeconomic}.
Lexical ambiguity of words can be assessed by comparing domain-specific meanings of words with their common meaning, e.g., by using BERT word embeddings with the Bradley-Terry statistical model~\cite{LIU2022103000}.
On the \textbf{document level}, the Flesch Reading Ease (FRE)~\cite{flesch1948new} is a commonly used metric for readability~\cite{becher2021law}. However, FRE does not capture readability holistically~\cite{crossley2017predicting}. 
Modern NLP-based approaches like TAASSC 2.0~\cite{kyle2021comparison} provide metrics on surface indices, syntactic features, and semantics. 
Comparing the consistency and logic of semantics between sentences allows coherence metrics, e.g., DiscoScore~\cite{zhao-etal-2023-discoscore}, which measure thematic focus frequency and sentence connectivity. 
Finally, metrics on the \textbf{structural level} consider that a better-structured text improves the understanding~\cite{cocklin1984factors}. Assessing structure separately is recommended~\cite{power2003document}.

\subsubsection{Bias Metrics}
\label{subsec:represntl}
The identification of biases typically needs a list of \textbf{descriptor terms}. 
Such descriptor terms are word lists in specific demographic axes that capture the different groups that could be misrepresented. For example, HolisticBias~\cite{smith2022m} includes 600 descriptor terms across 13 different demographic axes such as ability, age, body type, nationality, politics, etc. Descriptor terms can be generated automatically, e.g., for assessing gender bias on legal documents~\cite{baker-gillis-2021-sexism}.
Assessing semantics has become the standard practice in literature for assessing representational fairness~\cite{caliskan2017semantics,garg2018,Rice2019,baker-gillis-2021-sexism,schroder2021evaluating,Gumusel2022}. To measure a bias, the text in question is transformed into \textbf{word embeddings}, i.e., a numerical, vectorized representation of the words used. Classical approaches measure \textit{cosine similarity} between the descriptor terms, utilizing word embeddings to detect biases~\cite{garg2018,Gumusel2022, caliskan2017semantics, schroder2021evaluating}. A popular metric is WEAT~\cite{caliskan2017semantics}, improved in its robustness to SAME~\cite{schroder2021evaluating}.

Other approaches are based on text features (e.g., using all-caps words or bold formatting) or semantic features (e.g., sarcasm or insinuation of dishonesty or deceitfulness). For example, see \cite{risch2021overview}, a solution for identifying toxic engaging, and fact-claiming comments. This, however, can be biased in itself and needs bias mitigation~\cite{garg2023handling}.

\subsubsection{Morality and Ethics Classification}
\label{subsec:moralityasses}
Ethics might question the general morality of processes~\cite{wallimann2021integrated}. A judgment could be derived following a set of principles~/~virtues like autonomy, nonmaleficence, beneficence, and justice~\cite{beauchamp2001principles}.
However, research on classifying morality or ethics focuses on moral stance prediction~\cite{pavan2020morality}. 

Modern Large Language Models (LLM), like Mixtral-8x7B~\cite{jiang2024mixtral} or GPT4turbo~\cite{OpenAI2023}, show some cross-domain understanding~\cite{hendrycks2020measuring}, capabilities of reasoning~\cite{bubeck2023sparks} and some understanding of the concepts of morality and ethics~\cite{pock2023llms}. 
To evaluate ethics and morality with them, modern prompt engineering approaches~\cite{nori2023can} are required.

\subsection{Evaluation of Privacy Policies}
\label{subsec:study}
Fairness in privacy policies has been manually assessed in the mobile health domain~\cite{benjumea2020assessment}. A rather narrow definition of fairness based on items of the 14 content requirements stated in Article 13 of the GDPR is embraced. We are unaware of an automated approach to assessing fairness of privacy policies independent of the application domain. 
Research shows that privacy policies tend to be lengthy and use inaccessible language~\cite{transparency,becher2021law}. For these reasons, approaches summarize or tag key aspects of privacy policies with NLP and machine learning~\cite{harkous2018polisis,nokhbeh2020privacycheck, tesfay2018privacyguide}. The completeness and compliance of privacy policies regarding the GDPR~\cite{contissa2018claudette,amaral2021ai, 10.1145/3184558.3186969, sanchez2021automatic,vanezi2021complicy,xiang2023policychecker,torre2020ai,elluri2021bert} can be automatically assessed. 
Closest to our work is Claudette~\cite{contissa2018claudette}, which assesses GDPR compliance based on completeness of provided information, substantive compliance of a policy, and clarity of expression. Assessing fairness of policies goes beyond assessing their compliance with law. Our assessment of informational fairness is also concerned with completeness of provided information, assessing readability in our approach also tries to identify vague language but, further, addresses issues like readability for protected groups. Our approach assesses representational fairness, which Claudette does not cover, and ethics, which Claudette partially covers with substantive compliance.
The most promising approach for completeness of a policy~\cite{amaral2021ai} first identifies informational requirements involving 56 metadata types relevant to the completeness of a policy. Based on that, it identifies 23 criteria to check for policies. The completeness assessment via Machine Learning (ML) and NLP is based on semantic similarity and word embeddings.
Other approaches provide criteria or templates for more user-friendly and understandable policies~\cite{renaud2018make,feng2021design}.

To the best of our knowledge, representational fairness has not yet been assessed for privacy policies. However, representational fairness has been successfully assessed in loosely related fields of the legal domain~\cite{Rice2019,baker-gillis-2021-sexism,Gumusel2022} or more broadly on textual data~\cite{garg2018,caliskan2017semantics,schroder2021evaluating}. 

Assessing morality and ethics requires looking into what policies enforce. Judging ethics is a field that is typically addressed by ethics councils. Privacy policies have been found to obfuscate unethical data handling practices and use persuasive language~\cite{pollach2005typology}. However, the criteria of the GDPR can be used as a first step to evaluate the risks of a privacy policy. For example, policies have been found to lack specificity regarding data use practices~\cite{Zaeem2020} and lack protection of minors as well as a clear communication of changes made to their policy~\cite{contissa2018claudette}. 
A prominent approach to identifying risk is to use a Support Vector Machine to sentence- or paragraph-wise quantify the degree of policy compliance with the data protection goals~\cite{contissa2018claudette,sanchez2021automatic}. 
Most approaches do not make their data and code available.

\section{Problem Statement}
\label{sec:statement}

In this section, we explore the legal foundation of fairness in privacy policies, we derive three dimensions of fairness, and we propose a working definition for fair privacy policies.

\subsection{Legal Foundation}
Equal treatment is a fundamental human right, as declared in anti-discrimination principles in the Universal Declaration of Human Rights (UDHR)~\cite{UDHR}, the European Convention on Human Rights (ECHR)~\cite{ECHR}, and respective regulations in national law, e.g., the US~\cite{edenberg2023disambiguating}. 
Art. 12 of the UDHR and Art. 8 of the ECHR emphasize the right to privacy. The Privacy Guidelines~\cite{oecdprivacyguidelines} of the Organization for Economic Co-operation and Development (OECD) provide a framework for data privacy protection that is widely used as a reference for legislation. The guidelines require that \enquote{data should be obtained by lawful and fair means}, and implicitly reference various fairness concepts to establish a balanced approach to privacy protection. 
The OECD Privacy Guidelines have been widely adopted. For example, the 21 member states of the Asia-Pacific Economic Cooperative mirror the Guidelines~\cite{APECPrivacyFramework2015}. In the USA and Canada, the OECD Guidelines are the basis of the Generally Accepted Privacy Principles (GAPP)~\cite{AICPACICA}. The GAPP are \enquote{based on internationally known fair information practices}~\cite{AICPACICA}.
In the European Union, the GDPR~\cite{eu2016regulation} implements these guidelines into EU law. Chap.~2 GDPR specifies the openness and participation principles, which address informational fairness. Chap.~5 GDPR requires \enquote{ensuring that there is no unfair discrimination}.  Art. 5(1) GDPR enforces fairness as a key principle for privacy policies~\cite{clifford2018data,malgieri2020concept}. 

\subsection{Dimensions of Fairness}
Legal foundations motivate our three dimensions of fairness:

\textbf{Representational Fairness:} Art. 1 UHDR states that equal treatment is a fundamental human right. 
Art. 2 UHDR and Art. 14 ECHR explicitly forbids \enquote{\textit{discrimination} on any ground}.

\textbf{Informational Fairness:} Art. 
13 and 14 GDPR enforce \textit{completeness} of the information provided in a privacy policy on how it handles its users' data and what rights its users have. 
Art. 12(1) GDPR aims at making privacy policies \textit{readable} by requesting \enquote{concise, transparent, intelligible and easily accessible form, using clear and plain language}. Recital 39 requests \enquote{any information and communication of the processing of those personal data to be easily accessible and easy to understand}. 

\textbf{Ethics and Morality:} Art. 5(1) point (a), 6(2)-(3) GDPR and Re\-ci\-tal 71 states \textit{fairness} as a fundamental principle, grounded in ethics. The GDPR strives to avoid potential harm to data subjects through safeguards, and represents procedural fairness as a central element of its fairness conception~\cite{malgieri2020concept}. This means users should be protected from \textit{vulnerabilities and risk}. Interests between the data subject and service provider should be \textit{fairly balanced}~\cite{clifford2018data,malgieri2020concept}.

In summary, a privacy policy introduces (among other aspects) fairness into the interactions between the data controller and data subject. It cannot deliver fairness, if it is unfairly formulated. Therefore, our working definition of fairness in privacy policies is: 

\textit{\textbf{A fair privacy policy complies with informational fairness and representational fairness, as well as ethics and morality.}}

We are open to considering more dimensions of fairness in the future. 
We want to highlight that our three fairness dimensions are just motivated by the legal foundation and go beyond legal compliance.
We propose approaches to assess indicators for each dimension, we show the results of preliminary experiments, and we discuss how this brings forward fairness in privacy policies.

\section{Informational Fairness}
\label{sec:informational}

Our assessment of informational fairness borrows from the approaches described in Section~\ref{subsec:readbl}. 
\subsection{Approach}
Informational fairness in privacy policies is related to completeness of a privacy policy and comprehensibility and readability of vocabulary, sentences, and document structure for protected groups.

\textbf{Completeness:}
To get an estimate of the completeness of privacy policies, we propose to use supervised learning, as described in~\cite{amaral2021ai}. We want to identify the relevant 56 metadata types and then assess the embedding space to judge completeness.

\textbf{Fairness on word level:}
Anglicisms, complex words, tech jargon, etc. induce a socioeconomic, nationality, ableist, and age-related bias, causing unfairness due to discrimination. 
We propose lexical filtering, either with an English language dictionary, a word-to-word translation library, or a customized dictionary, to detect tech jargon and anglicisms. 
A frequency-based dictionary~\cite{Schaefer2015b, SchaeferBildhauer2012full} allows us to estimate the proportion of words in a policy, that are not in common usage with a threshold-based approach. 
We can estimate the use of complex or ambiguous words, when we use a translation service to translate words into another language and back, and measure the proportion of words that remain unchanged. 
However, the translation service might be affected by biases~\cite{savoldi2021gender} and linguistic limitations, e.g., when translating a gendered language to a gender-neutral one and back. 
We also propose the Bradley-Terry statistical model with embeddings to assess lexical ambiguity~\cite{LIU2022103000}.

\textbf{Fairness on sentence level}:
Incomprehensible policies lead to discrimination and unfairness. 
We propose to estimate the readability of the sentences using statistical text metrics and existing linguistic measures. 
A straightforward text metric is the number of words in the policy. A higher number indicates less readable policies. The Flesch Reading Ease (FRE)~\cite{flesch1948new} and the TAASSC 2.0~\cite{kyle2021comparison} are well-recognized linguistic metrics for readability. 
The coherence of sentences can be assessed with the DiscoScore metric~\cite{zhao-etal-2023-discoscore}.

\textbf{Inclusive document structure:}
An inclusive policy structure mitigates socioeconomic, nationality, and ableist bias and improves fairness. Again, we propose statistical and linguistic measures. An inclusive policy is indicated by short section headings or lengths per paragraph, or a shallow split of the policy into sections (structural depth).
To assess how well the headings semantically fit with the whole section text of a policy, embeddings can be used~\cite{kenter2015short, han2021survey}.

\subsection{Preliminary Results}
 
To provide evidence, that our work-in-progress indeed allows to assess informational fairness in privacy policies automatically, we have conducted a series of prototypical experiments regarding wording, sentences, and document structure. We leave experiments, that require a high implementation effort, to future work.

As a test case, we use the German Top-100 most visited web shops. Our data set contains the canonical form of 618 German privacy policies over eight years, starting in 2016. For acquisition, preprocessing, cleansing, and a description of the data set, see~\cite{transparency}. We conducted all experiments in German, as this is the language of the data set, but translated the results to English for this paper. 

Concerning fairness on word level, we exemplarily selected eight policies, one for each year in the data set, and conducted lexical filtering. A simple dictionary-based approach to filtering for potential anglicisms produces many false positives. Many German stop words in NLTK~\cite{bird2009natural} overlap with English words from SpaCy's~\cite{spacy2} English dictionary and can be removed. We then used a word-to-word translator~\cite{choe2020word2word} with a smaller vocabulary than SpaCy or NLTK, to narrow down our list of anglicisms, and measured the length of this list. In our eight policies, we found 26 anglicisms on average, which impairs informational fairness by discriminating against demographics, where anglicisms are not commonly used. 

For assessing fairness at sentence level, we used the entire data set of 618 policies. 
We measured an average FRE of 37, with FRE=14 as the worst and FRE=88 as the best value. In comparison, an FRE 30 or below requires the reading competence of an academic. A fair, non-discriminating policy should read an FRE of 60 and above.  

\begin{table}[ht]
\centering
\caption{\label{tab:1} Overview of surface measures}
\begin{tabular}{p{2.8 cm}|p{3.8cm}}
\textbf{Measure}	& \textbf{Average or Value}	\\
\hline
words / policy	& 	4809.59\\
\hline
paragraphs / policy & 149.63\\
\hline
words / paragraph	& 32.14	\\
\hline
headings / policy	& 	33.82 (6.15\% have no heading)\\
\hline
heading types	& 	2.85\\
\hline
words / heading	& 	4.83\\
\hline
lists	& 	91.59 \% of policies\\
\hline
other text formatting	& 	strong: 71.52 \% of policies\par italics: 15.37 \% of policies\\

\end{tabular}
\end{table}
Table \ref{tab:1} provides statistics on the document structure of all 618 policies. 
The average policy is very long. A dyslexic, who needs $\approx$2 minutes for 250 words~\cite{musch2011schnell,martelli2014bridging}, would need to spend almost 165 minutes reading the longest policy. The average reading time per policy for an average reader~\cite{musch2011schnell} would be just under 20 minutes.

On average, a policy has 150 paragraphs with 32 words each. It contains 34 headings on 3 levels, with an average heading length of 5 words. 
92\% of the policies make use of lists, 72\% contain strong formatting, and use 15\% italic. However, 6\% of the policies are not structured by headings at all. This leaves a mixed impression regarding informational fairness in privacy policies.

\subsection{Discussion}

Text statistics, FRE and, to some extent, formatting have already been measured for privacy policies aiming at transparency~\cite{transparency}. 
However, existing work did not use this information to assess key aspects of fairness. 
While informational fairness has previously been assessed for the predictions of machine learning models~\cite{Schoeffer2022}, privacy policies are a different problem. Privacy policies are intended to balance the information asymmetry between provider and user (completeness), and add the aspect of potential discrimination due to a lack of inclusiveness in the presentation of information. 
Most approaches to privacy policies from the related work (cf. Sec.~\ref{subsec:study}) only focus on completeness of information. Our preliminary results address different issues and allow for no comparison.

\section{Representational Fairness}
\label{sec:representational}

Our approach regarding representational fairness uses the methods outlined in Section~\ref{subsec:represntl} to assess the ability-related, socioeconomic, age-related, political, nationality-related, gender-related, sexuality-related, and cultural social~/~demographic axes.

\subsection{Approach}
Assessing representational fairness requires descriptor terms that capture various demographics (cf. Sec.~\ref{sec:related}). Based on the descriptor terms, a quantitative and semantic assessment can be carried out.

\textbf{Fairness descriptor terms:} 
We propose to carefully select existing descriptor terms, e.g., for gender~\cite{muller2023gender}, or via translation service from other languages~\cite{smith2022m}. Finally, we propose to use a multilingual LLM to filter the resulting set, generate contextualized translations, or automatically generate policy-specific descriptor terms.

\textbf{Quantitative and semantic fairness assessment:} 
The descriptor terms enable us to estimate how often different groups are represented in a policy. Semantically, we propose to use the descriptor terms to assess the invariance of sentiment toward changed protected group membership by replacing existing entities with changed ones of varying protected groups. 
Finally, we propose to leverage the SAME metric~\cite{schroder2021evaluating} for bias detection. 
We want to adapt approaches for toxicity detection~\cite{risch2021overview}, and debiasing the detection itself~\cite{garg2023handling}.

\subsection{Preliminary Results}
Again, our test case is based on German language privacy policies from the Top-100 German web shops. With this series of experiments, we evaluated the entire data set of 618 policies. 
We translated the descriptor terms from HolisticBias~\cite{smith2022m} to German with deep-translator~\cite{deep-translator} utilizing the Google Translator API, and we manually removed homonyms and other artifacts from the translation. 
We appended terms from the categories \enquote{nationality} and \enquote{political} that are specific for a German context, and we integrated German gender descriptor terms~\cite{muller2023gender}. We perform a quantitative assessment.

We found that the policies do not take into account reading-impairing disabilities. No specific disability from our descriptor terms is mentioned even once. Some policies address the reader informally. We find terminology that is typically used to address youth readers mentioned 13325 times overall. 
The vocabulary is rather gender-neutral (13 male occurrences, 130 female occurrences, 6660 gender-neutral occurrences) based on the German gender descriptor terms. However, policies lack appropriate gendering. For instance, we found the German word for "user" 3057 times not gendered. 
This could be problematic for the inclusiveness of poor readers, age-related demographics, women, or people who identify as non-binary. 

We observed difficulties when assessing age-related representational bias, because \enquote{old} and similar descriptor terms are frequently but differently used in privacy policies. This needs a refinement of the word list and assessing the semantic space. The quantitative assessment suggests a nationality bias toward the US, Germany, and generally European countries. This also needs to be checked by assessing semantics, as it may just be related to information about the stakeholders' locations. We could not find quantitative representational bias regarding political, cultural, or sexuality-related demographics.

\subsection{Discussion}
To the best of our knowledge, representational fairness has not yet been assessed for privacy policies. We have only investigated quantitatively as a first step. For more nuanced and accurate findings, we want to assess semantics as a next step. This addresses false positives due to homonyms and has been used effectively in recent research (cf. Sec. ~\ref{subsec:represntl}).

\section{Ethics and Morality}
\label{sec:ethics}
This assessment uses approaches from Section~\ref{subsec:moralityasses} to screen the processes, rights, and obligations declared in a privacy policy. 

\subsection{Approach}
We propose an LLM to assess to which vulnerabilities or risks a policy exposes a data subject, how proportional those are, and which general ethics issues exist. 

\textbf{Vulnerabilities and Proportionality:}
We propose to train a specific vulnerability classifier, similar to~\cite{sanchez2021automatic}. Based on that, the proportionality of a policy depends on the business activities of the data holder. The assessment of proportionality requires weighing the difficult-to-quantify reasons for a case-specific process against its induced vulnerabilities. This raises challenges for typical ML classification models, so we suspect an ethicist's judgment to be superior (cf. Section~\ref{sec:related}). That's why we propose utilizing an LLM as a new approach to be tested. We use the classification result of the vulnerability classifier to prompt the LLM.

\textbf{General Ethics:}
We also suggest testing general ethics issues with an LLM, using a broad framing of the prompt. 
Because this is an explorative approach, we propose to request the LLM to state the criteria of its assessment first, and to quantify them on a five-point Likert scale in a subsequent step. 
We know from prompt engineering~\cite{nori2023can} that this can be achieved by using the entire policy as context and a task description in the prompt, and by asking the LLM to reflect its result.
As a supplement, we suggest approaches for sentiment analysis~\cite{guhr-EtAl:2020:LREC} on the LLM outputs, to get an approximate overall proxy for the ethics of a policy.

\subsection{Preliminary Results}
Again, we use the German language policies from the German Top-100 web shops. 
For the time being, we focus on general morality and leave aside the other aspects. 
We created a prompt with a broad scope (see Figure~\ref{fig:prompt}), and tested it with GPT4turbo~\cite{OpenAI2023} on eight policies, one for each year in the data set. We asked the LLM not only to make a qualitative assessment, but also to rate it on a five-point Likert scale. Due to the lack of space, we only report the Likert ratings (Table~\ref{tab:2}).

As the first column of Table~\ref{tab:2} shows, we can aggregate the LLM assessments into 12 distinct criteria for the ethics of privacy policies. 
As desired, the criteria capture general ethical requirements on a broad scope and go beyond the GDPR, e.g., regarding the protection of minors. 
An LLM answers each prompt with a new internal seed, comparable to an expert asked to make an assessment from a different point of view. 
For this reason, we let the LLM evaluate each prompt in five runs. The second and third columns of Table~\ref{tab:2} show in how many runs and policies the criteria appear. Some criteria just appeared in one of the runs for a policy.
While integrating a fixed set of criteria into the prompt would enable a standardized assessment, running the LLM multiple times results in different aspects being considered, which has some merit for exploration.
The fourth column shows the Likert-scores between 1~(worst) and 5 (best). Our most alarming findings are the usage of data, the (international) transmission, and sharing of data, as well as communication and changes.

\begin{figure}[htb]
    \centering
\fbox{\parbox{.9\linewidth}{\small 
\em
\textbf{System prompt:} You are a critically scrutinizing, experienced expert in an ethics council on issues around fairness and justice and are committed to consumer protection for fairness. You have 20 years of experience in ethical issues around consumer protection, data protection, cybersecurity and related fairness aspects. You have already provided precise expert opinions in many court proceedings.
Your task is to uncover aspects in privacy policies that are ethically questionable. You must clearly explain your thoughts and how you came to your conclusion. Explain step by step! Also disclose the ethical and moral principles on which you base your judgment. In particular, point out potential for improvement or risks to users. 
IMPORTANT: Also rate the privacy policy for each of your criteria on a 5-point Likert scale. Check whether the following privacy policy is fair and ethical towards its users.
The user wants a critical review and an assessment of vulnerabilities in privacy policies.

\textbf{User prompt:} The privacy policy: [Privacy Policy]
}}
    \caption{Prompt template (Square brackets replaced with privacy policy; German template translated to English)}  
    \label{fig:prompt}
\end{figure}

\begin{table}[ht]
\centering
\caption{\label{tab:2} Overview of LLM ethics assessments}
\begin{tabular}{p{4.1 cm}|p{0.8 cm}|p{1 cm}|p{0.9 cm}}
\textbf{Criteria} & \textbf{\#} \par \textbf{Runs} & \textbf{\#} \par \textbf{Policies}	& \textbf{Average} \par \textbf{Score}	\\
\hline
\textbf{transparency} (precise, complete, comprehensible) &  38 &  8  &  3.20  \\
\hline
\textbf{data subject control and autonomy} (user rights, freedom of choice) &  39 &  8  & 3.35   \\
\hline
\textbf{data minimalization and purpose binding} & 26  & 8   &  3.06  \\
\hline
\textbf{data usage} (surveillance concerns, automated decisions) &  21 &  8  &  2.71  \\
\hline
\textbf{data storage and deletion} & 8  &  4  & 3.38   \\
\hline
\textbf{data protection and security} (avoiding misuse or leakage) & 30  &  8  & 3.38   \\
\hline
\textbf{(international) data}\par \textbf{transmission and sharing} (third-parties, where is what transferred) & 27  &  8  & 2.74   \\
\hline
\textbf{compliance with data protection regulation} (GDPR, relevant countries data protection regulation and standards) &  9 &  7  &  3.56  \\
\hline
\textbf{communication and changes} (timeliness, changes, risks) & 8  & 5   &  2.63  \\
\hline
\textbf{protection of minors} &  4 & 2   &  3.50  \\
\hline
\textbf{fairness} (no discrimination, fair use) & 7  &  4  & 3.14   \\
\hline
\textbf{assurance, accountability and governance} (compliance, reliability, assurance, ability
to react) & 9  &  6  &  3.22  \\
\end{tabular}
\end{table}

\subsection{Discussion}
Existing work has not automated ethics assessment so far, i.e., an LLM for ethics evaluation of privacy policies is a novel approach. Our findings are in line with related work (cf. Sec. ~\ref{subsec:study}) regarding problematic data use.
We found the GDPR principles, which are hard ethics, represented in the ethics criteria, which we identified automatically with GPT4turbo. 
As we exemplarily investigated eight policies with a multilingual LLM, our list of criteria is just a starting point. There may be further criteria worth considering by utilizing a German LLM, more sophisticated prompting and a large and diverse set of policies. 

\section{Use Cases and Applications}
\label{sec:uses}

In this section, we explain why our approach has the potential to improve the fairness of privacy policies. In particular, we see two prominent use cases in the areas of \textit{analyses} and \textit{writing support}. 

Our approach enables researchers to automatically scan a large corpus of privacy policies for such issues. This allows to obtain an overview of how widespread which kind of fairness issue is in privacy policies. Due to the ties between fairness and the GDPR, this might also help authorities to single out privacy policies that are problematic from a legal point of view. Our data set allows to analyze how such issues have developed or subsided over time. This provides an opportunity for researchers to investigate whether changes in legislation or major privacy events have led to fairer or less fair policies. 
Writing a fair policy is challenging. In the case of AI applications, for example, it is difficult even for experts to assess the personal risks associated with the use of personal data. Our approach allows to automatically identify aspects, that are associated with various kinds of bias, discrimination, etc. It might be useful to integrate this either into existing generators for privacy policies or to develop a stand-alone tool, to support data protection officers in writing complex, but fair privacy policies.
Also, our work can be used as a foundation for data assessment for legal NLP models trained on privacy policies.

\section{Conclusion}
\label{sec:conclusion}

Assessing privacy policies regarding their fairness is an important, but yet unresolved issue.
In this paper, we suggest three dimensions of fairness that should be investigated: informational fairness, representational fairness, and ethics and morality. We base our problem understanding in literature on fairness, privacy policies, and fairness issues in legal text processing. Furthermore, we propose assessment procedures for all three fairness dimensions, utilizing NLP and linguistic analysis. 

We implemented and tested a small subset of those procedures, with promising results. Our next step is to implement a holistic system of assessments, as suggested in our approach. Beyond that, we seek a close collaboration with ethical and legal experts to identify further relevant criteria that can be used to iterate, find out what is needed for a more complete assessment procedure, and how our method could be used by ethicists and jurists as tool support.
Our contribution is important as we advance the understanding of fairness in privacy policies. Our approach may help in providing data subjects with some much-needed transparency and in avoiding discrimination or unethical practices.


\bibliographystyle{ACM-Reference-Format}

\balance
\bibliography{literature}

\end{document}